\documentclass[ijoc,nonblindrev]{informs3} % current default for manuscript submission

\OneAndAHalfSpacedXII % current default line spacing
\usepackage{natbib}
 \bibpunct[, ]{(}{)}{,}{a}{}{,}%
 \def\bibfont{\small}%
\TheoremsNumberedThrough     % Preferred (Theorem 1, Lemma 1, Theorem 2)
\EquationsNumberedThrough    % Default: (1), (2), ...
\begin{document}
%%%%%%%%%%%%%%%%

%\setcounter{page}{1}
%\DOI{10.1287/xxxx.0000.0000}%

\RUNAUTHOR{Hossein Ghorban and Hesaam}

\RUNTITLE{Product Portfolio Management in Competitive Environments}

% \TITLE{Bundling Information Goods of Decreasing Value}
\TITLE{Product Portfolio Management in Competitive Environments}

\ARTICLEAUTHORS{%
\AUTHOR{Samira Hossein Ghorban, Bardyaa Hesaam}
\AFF{School of Computer Science, Institute for Research in Fundamental Sciences (IPM), P.O.Box 19395 - 5746, Tehran,  Iran \EMAIL{s.hosseinghorban@ipm.ir}, \EMAIL{bardyaa@ipm.ir}}
%\AUTHOR{Author2}
%\AFF{Author2 affiliation, \EMAIL{}, \URL{}}
} % end of the block

\ABSTRACT{%
Product diversity is one of the prominent factors for customers' satisfaction, while from the firms' perspective,
the additional engineering costs required for product diversity should not exceed the acquired profits from the increase in their  market share. Thus, one of the critical decision-making tasks for companies is the selection of an optimal mix of products, namely  product portfolio management (PPM). Traditional studies on PPM problem have paid relatively less attention to the actions of other competitors. In this paper, we study PPM problem in a competitive environment where each firm's objective is to maximize its expected shared surplus. We model the competition with an $n$-player game that optimal product portfolios are driven from its Nash equilibrium. Utility functions are determined by the expected value of the shared surplus. We analyze the strategic behavior of firms to determine their optimal  product portfolios. % Enter your abstract
}%

%\KEYWORDS{deterministic inventory theory; infinite linear programming duality; 
%  existence of optimal policies; semi-Markov decision process; cyclic schedule}


\maketitle
%%%%%%%%%%%%%%%%%%%%%%%%%%%%%%%%%%%%%%%%%%%%%%%%%%%%%%%%%%%%%%%%%%%%%%

% Samples of sectioning (and labeling) in IJOC
% NOTE: (1) \section and \subsection do NOT end with a period
%       (2) \subsubsection and lower need end punctuation
%       (3) capitalization is as shown (title style).
%
%\section{Introduction.}\label{intro} %%1.
%\subsection{Duality and the Classical EOQ Problem.}\label{class-EOQ} %% 1.1.
%\subsection{Outline.}\label{outline1} %% 1.2.
%\subsubsection{Cyclic Schedules for the General Deterministic SMDP.}
%  \label{cyclic-schedules} %% 1.2.1
%\section{Problem Description.}\label{problemdescription} %% 2.

% Text of your paper here

\section{Introduction}
Portfolio management is strategic decision-making or strategic planning to forecast how firms should spend their scare engineering, operation resources, and marketing resources to maximize their objective functions which is categorized as resource allocation problems.
It focuses on having the right balance between number of projects and available resources or capabilities~\cite{cooper1999new}.
However, an organization must optimize its product diversity to  increase  revenues~\cite{lancaster1990economics}.
There are two methods for a business to succeed new products: \textit{doing projects right}, and \textit{doing the right projects}~\cite{cooper2000new}.
Hence, an important decision-making for the firms is to offer the \textit{right} product variety to the target market instead of creating various products in relation to anticipate all needs of customers~\cite{jiao2007heuristic}.

Portfolio management has been extensively studied with  different approaches.
A linear programming method is applied for {R}\&{D} project selection~\cite{jackson1983decision}.
%deta an experiment reported in~\cite{baker1974r}, the author comments on the major obstacle of the amount of date which are required. 
Financial models and financial indexes, 	 probabilistic financial models, options pricing theory,
strategic approaches,  scoring models and checklists,
analytical hierarchy approaches,
behavioral approaches,
mapping approaches or bubble diagrams are different methods for selecting a new portfolio~\cite{cooper1999new}.
While there are several approaches allowing companies to increase  R\&D productivity, their implementations impose  new challenges for product portfolio management. In~\cite{cooper2020new}, new solutions are proposed to deal with these emerging challenges. The problem of product portfolio optimization is analyzed by population dynamic approach in which 
company's product portfolio is considered  as a product population~\cite{wang2021collaborative}. The authors analyze  the product population’s growth balance  by the logistic model.

It also considers how the product population’s scale and structure can be continuously optimized in a way that balances enterprise output, product synergy, and resource allocation.

The conjoint-based approach to optimal product portfolio problem ends in integer optimization problems which are   NP-hard problems~\cite{nair1995near}.
For instance, the problem of  designing a new product line in order to maximize its surplus which is determined based on customer preference is NP-hard~\cite{kohli1989optimal}.
In~\cite{belloni2008optimizing}, efficient methods have been devised for the customer preference perspective.  
Furthermore, product portfolio decisions have been examined with a focus on the engineering implications, the cost and complexity of actions among multiple products~\cite{simpson2004product}.

Furthermore, Jiao and Zhang~\cite{jiao2005product} investigated product portfolio management with the view of customer-engineering interaction and established a
maximizing shared-surplus model for PPM problem in the absent of competition view point where the detail of produce design has been included.
They formulate the problem as an integer programming which is an NP-hard problem and also a heuristic genetic algorithm  is applied to solve the relaxed version of the integer linear programming problem.
However, diverse analyses have been applied to examine PPM  problems.
One of them is the concept of Nash equilibrium has been employed to model competitive reactions in produce design
\cite{choi1993game} and product line design \cite{kuzmanovic2012approach}.

The new entrant firm into a competitive market has been
studied in~\cite{choi1990product,steiner2010stackelberg,liu2015stackelberg}.
The new entrant firm has more resources and pre-experience on the rivals' behavior.
Hence, it is can be expected to become a leader against the other firms in the market~\cite{choi1990product}.
The Stackelberg (leader-followers) game has been applied to find out an optimal for a single product design~\cite{steiner2010stackelberg} and product portfolio~\cite{liu2015stackelberg}. 
The objective of~\cite{liu2015stackelberg} is to maximize  the expected
shared surplus of the new market entrant. 
Steiner formulates optimal product design problem based on the perspective of a
profit-maximizing new entrant (the leader) who wants to launch a brand onto an
existing product market and acts with foresight by anticipating price-design reactions
of the incumbent firms (the Nash followers)~\cite{steiner2010stackelberg}
while Liu et al.'s concern is to maximize  the expected shared surplus of the new market entrant.

The competitive interactions of two firms to find an optimal product portfolio has been modeled by a non-cooperative complete information game~\cite{sadeghi2011game}. In this game, utility functions for firms are determined   based on the customer-engineering interaction model which is proposed in~\cite{jiao2005product} and Nash equilibrium of this
game is calculated for just one  numerical example where two firms with four different products compete in a market with 3 segments.
Product line design as an instance of PPM problem has been studied with the game theoretical approach.  In~\cite{liu2017product}, the competitive interactions of firms to design product line has been formulated with the Stackelberg model and for an industrial case of cell phones and the analysis of finding an equilibrium of this  model has been implemented. 
Moreover, the competition environment between two firms from the viewpoint of the product cost and customer satisfaction has been modeled by a Bayesian game~\cite{yang2019configuration}.

John Nash introduced a solution concept for strategic-form games, called a Nash Equilibrium (NE)~\cite{nash1951non}.
He proved that every finite strategic-form game has a mixed strategy equilibrium. He presented
two existence proofs: the first one was based on Kakutani’s Fixed Point Theorem~\cite{nash1950equilibrium}, and the
second one was based on Brouwer’s Fixed Point Theorem~\cite{nash1951non}. 

C. Papadimitriou~\cite{papadimitriou1994complexity} proposed the complexity class PPAD (Polynomial Parity Argument in a Directed graph) which is the class of all search problems that can be polynomially reduced to the  END OF THE LINE problem: 
given two circuits $S$ and $P$, each with $n$ input bits and $n$ output bits, such that 
$P(0^n)=0^n=S(0^n)$, find an input $x \in {\{0,1\}}^n$ such that 
$P(S(x))\neq x$ or $S(P(x)) \neq x \neq 0^n$.
PPAD class contains several important problems that are suspected to be hard~\cite{daskalakis2009complexity}.
Moreover, it is shown that finding a NE for a finite strategic-form game is PPAD-complete~\cite{daskalakis2009complexity}.

In this paper, we study an $n$-agent game which is a generalization of 2-agent game was developed in~\cite{sadeghi2011game} 
to model a competitive market with different segments for product portfolio management. 
The self-interest of each firm is to maximize its total expected shared surplus which is formulated in~\cite{jiao2005product}.
Luxury brands  in which the central point of such markets concentrates on the rich consumers is one example of  a single market segmentation.
In~\cite{sadeghi2011game}, an example of two firm who compete in a single market segmentation has been examined. We analyze the process of finding a mixed Nash equilibrium for these game in a single market segmentation when all strategies in NE are inner points of strategy spaces which we call it an \textit{interior Nash equilibrium}.

This paper is organized as follows.
In Section~\ref{Problem Description}, the PPM problem in competitive environment is formulated with an $n$-agent game. An analysis for finding an interior Nash equilibrium for a single market segmentation is presented in Section~\ref{NEAnalysis}.
Finally, the conclusion are presented in Section \ref{conclusion} with a number of areas for future works.
%----------------------------------------------------------------------
%----------------------------------------------------------------------------
\section{The Model} \label{Problem Description}
Suppose that $Z$ is a potential product set for a market which has multiple market segments, represented by $\{G_j : 1\leq j \leq m\}$ such that each segment $G_j$ contains homogeneous customers with total demand $Q_j$. We enumerate the product set $Z$ with $\{1,\ldots, \rho\}$ where $\rho =|Z|$.
Assume $n$ firms, $I=\{1,\ldots,n\}$, compete in the market.
Let $Z_i \subseteq Z$ be the set of products which are technically possible for  firm $i\in I$.
Thus, a product portfolio for  firm $i$ is a subset of $\varLambda_i \subseteq Z_i$ such that $\varLambda_i \ne \varnothing$. Hence, the pure strategy set for firm $i \in I$ which we denote it by $S_i$ is a set of non-empty subsets of $Z_i$. Consequently, a mixed strategy for this firm is a distribution on $S_i$.  

Suppose that $\hat\sigma_i : S_i\to [0, 1]$ is a mixed strategy for firm $i$. We show that it induces a distribution $\sigma_i$ on $Z_i$. Let
\[\sigma_i(p) = \sum_{\varLambda\in S_i\atop p\in \varLambda} \frac{\hat\sigma_i(\varLambda)}{|\varLambda|}.
\]
As each $\varLambda\in S_i$ appears as many times as its elements, it follows that 
$\sum_{p\in Z_i} \sigma_i(p) = 1,$
which in turn shows that $\sigma_i$ is a distribution on $Z_i$. On the other hand, if $\sigma_i$ is a distribution on $Z_i$ and $\varLambda\in S_i$, we may define
\[\hat\sigma_i(\varLambda) = \sum_{p\in\varLambda} \dfrac{\sigma_i(p)}{|\{\varGamma\in S_i : p\in\varGamma\}|},\]
which gives a distribution on $S_i$.

Every product has  certain engineering costs; so different firms may have different production costs due to different technologies. Thus, we denote the price of product $p \in Z$ produced by firm $i\in I$ in $G_j$, $1 \leq j \leq m$ by $\beta_{ijp}$.
Assume, for  firm $i$, customer preference for  product $p\in Z_i$ in $G_j$ is represented by respective utility, $u_{ijp}$.
Now, we model  the competition among $n$ firms as a game $G$   and call it a PPM game. 
\begin{definition}
	A PPM game $G$ is a game in strategic form with  the following structure
	\begin{enumerate}
		\item The set of firms, $I=\{1, \ldots , n \}$.
		\item The set of $m$ market segments $G_j$ where $ 1 \leq j \leq m$.
		\item Firm $i$'s strategy space 
		$$\varSigma_i=\big\{(\sigma_{i1},\ldots, \sigma_{i\rho})\in [0,1]^{\rho}\hspace{.1cm}:
		\hspace{.1cm} \sum_{p=1}^{\rho } \sigma_{ip}=1, \hspace{.1cm} p\notin Z_i \Rightarrow \sigma_{ip}=0\big\}.$$
		\item Firm $i$'s payoff for the strategy profile $(\sigma_i, \sigma_{-i})\in \varSigma_i \times \varSigma_{-i}$ is
		$$u_i(\sigma_i,\sigma_{-i})=\sum_{j=1}^{m} \sum_{p=1}^{\rho} \beta_{ijp} \cdot Q_j \cdot \frac{\exp(u_{ijp})\cdot \sigma_{ip} }{\sum_{r=1}^{n}\sum_{q=1}^{\rho}\exp(u_{rjq})\cdot  \sigma_{rq}\vphantom{A^{l^l}}} \cdot \sigma_{ip}.$$
		\end{enumerate}
\end{definition}
For convenience, let $e_{ijp}=\exp(u_{ijp})$ for each $i \in I$, $1 \leq j \leq m$ and $1 \leq p \leq \rho$.
\begin{remark}
	Let $(\sigma_i, \sigma_{-i})\in \varSigma_i \times \varSigma_{-i}$. 
	By the multinomial logit (MNL) model~\cite{mcfadden1977application},
	the probability that a customer chooses product $p$ produced by  firm $i$ in the segment market $G_j$ is equal to
	$$P_{ijp}(\sigma_i, \sigma_{-i}) 
	=\frac{e_{ijp}\cdot \sigma_{ip} }{\sum_{r=1}^{n}\sum_{q=1}^{\rho} e_{rjq}\cdot \sigma_{rq}}.
	$$
		for $i \in I$, $1\leq j \leq m$ and $1\leq p\leq \rho$.
\end{remark}

%-----------------------------------------------------------------------
\section{Equilibrium Analysis}\label{NEAnalysis}
John Nash introduced a solution concept for strategic games which is called Nash Equilibrium~\cite{nash1951non}.
He proved that every finite strategic-form game has a mixed strategy equilibrium. He presented
two existence proofs: the first one was based on Kakutani’s Fixed Point Theorem~\cite{nash1950equilibrium}, and the
second one was based on Brouwer’s Fixed Point Theorem~\cite{nash1951non}.  Since a PPM game is a finite game, it has a mixed Nash equilibrium. In this section, we explain how to find it when it is in $(0, 1)^{n\rho}$. 

Nash stated that a rational agent will only play the strategy which is a best response to the strategies actually taken by its opponents. To formalize the statement, let
\[\varDelta(S_i) = \{\sigma :  \mbox{$\sigma$ is a distribution on $S_i$}\},\]
where $S_i$ is a strategy set for agent $i$. 
The best response  of agent $i$ in a strategic-form game $G$ is a correspondence $BR_i : \varDelta(S_{-i}) \to \varDelta(S_i)$ given by
$$BR_i(\sigma_{-i})=\mathop{\mbox{arg max}}_{\sigma_i \in S_i} u_i(\sigma_i, \sigma_{-i})$$ for each $i$.
Clearly, strategy profile $(\sigma_1, \ldots, \sigma_n)  \in \prod_{i=1}^{n}\varDelta(S_i)$ is a  Nash equilibrium if and only if $\sigma_i  \in BR_i(\sigma_{-i})$ for each agent $i$.

Let $G$ be a PPM game, and suppose that \(\sigma^*\) is a Nash equilibrium. Then \(\sigma^*_i\) is a constrained (local-) maximum for \(u_i(\sigma_i, \sigma^*_{-i})\) subject to \(\sum_{p=1}^\rho \sigma_{ip} = 1\) for each \(1\leq i\leq n\); so if \(\sigma^*\) is an interior point of \([0,1]^{n\rho}\) then we may use the Lagrange multipliers method to find a subset of \(\mathbb{R}\) that must include \(\sigma^*\). To set things up, we first define 
\[g_i(\sigma_i)=\sum_{p=1}^\rho \sigma_{ip} - 1, \quad 1\leq i\leq n.\]
Then
\[\frac{\partial u_i}{\partial\sigma_i}(\sigma_i^*, \sigma_{-i}^*) \ \big\|\  \frac{\partial g_i}{\partial\sigma_i}(\sigma_i^*, \sigma_{-i}^*), \quad 1\leq i\leq n\]
where
\begin{align*}
\frac{\partial u_i}{\partial \sigma_i}(\sigma_i, \sigma_{-i})&=
\begin{bmatrix}
\dfrac{\partial u_i}{\partial \sigma_{i1}}(\sigma_i, \sigma_{-i}), \ldots, \dfrac{\partial u_i}{\partial \sigma_{i\rho}}(\sigma_i, \sigma_{-i})
\end{bmatrix}, \\
\frac{\partial g_i}{\partial \sigma_i}(\sigma_i, \sigma_{-i})&=
\begin{bmatrix}
\dfrac{\partial g_i}{\partial \sigma_{i1}}(\sigma_i, \sigma_{-i}), \ldots, \dfrac{\partial g_i}{\partial \sigma_{i\rho}}(\sigma_i, \sigma_{-i})
\end{bmatrix}.
\end{align*}
Hence, there are constants \(\lambda_1,\ldots,\lambda_n\in\mathbb{R}\) such that 
\[\frac{\partial u_i}{\partial\sigma_i}(\sigma_i^*, \sigma_{-i}^*) = \lambda_i\frac{\partial g_i}{\partial\sigma_i}(\sigma_i^*, \sigma_{-i}^*), \quad 1\leq i\leq n.\]
As \(\frac{\partial g_i}{\partial\sigma_i} = [1,\ldots,1]\), it follows that 
\[\frac{\partial u_i}{\partial\sigma_{is}}(\sigma_i^*, \sigma_{-i}^*) = \frac{\partial u_i}{\partial\sigma_{it}}(\sigma_i^*, \sigma_{-i}^*), \quad 1\leq i\leq n, \ 1\leq s,t\leq \rho.\]
Doing some calculations we get
\[\frac{\partial u_i}{\partial\sigma_{is}}(\sigma_i^*, \sigma_{-i}^*) = \sum_{j=1}^m \dfrac{2\beta_{ijs} Q_j e_{ijs} \sigma^*_{is}\left(\sum_{p=1}^\rho \sum_{r=1}^n e_{rjp}\sigma^*_{rp}\right)-\sum_{p=1}^\rho \beta_{ijp} Q_j e_{ijp} e_{ijs} {\sigma^*_{ip}}^2}{\left(\sum_{p=1}^\rho \sum_{r=1}^n e_{rjp}\sigma^*_{rp}\right)^2}.\]
From now on, we consider the case \(m=1\); so for all \(1\leq s,t\leq \rho\) we must have
\begin{align*}
2\beta_{i1s} Q_1 e_{i1s} \sigma^*_{is}\left(\sum_{p=1}^\rho \sum_{r=1}^n e_{r1p}\sigma^*_{rp}\right) &-\sum_{p=1}^\rho \beta_{i1p} Q_1 e_{ip} e_{i1s} {\sigma^*_{ip}}^2 \\
& = 2\beta_{i1t} Q_j e_{i1t} \sigma^*_{it}\left(\sum_{p=1}^\rho \sum_{r=1}^n e_{r1p}\sigma^*_{rp}\right)-\sum_{p=1}^\rho \beta_{i1p} Q_1 e_{i1p} e_{i1t} {\sigma^*_{ip}}^2.
\end{align*}
Doing some manipulations, we get
\[(e_{i1s}-e_{i1t})\sum_{p=1}^\rho \beta_{i1p} e_{i1p} {\sigma^*_{ip}}^2 = 2(\beta_{i1s} e_{i1s} \sigma^*_{is} - \beta_{i1t} e_{i1t} \sigma^*_{it}) \sum_{p=1}^\rho \sum_{r=1}^n e_{r1p}\sigma^*_{rp}\]
or
\[\frac{\beta_{i1s} e_{i1s} \sigma^*_{is} - \beta_{i1t} e_{i1t} \sigma^*_{it}}{e_{i1s}-e_{i1t}}
=\frac{\sum_{p=1}^\rho \beta_{i1p} e_{i1p} {\sigma^*_{ip}}^2}{2\sum_{p=1}^\rho \sum_{r=1}^n e_{r1p}\sigma^*_{rp}}.\]
As the right hand side does not depend on \(s\) or \(t\), we denote it by \(k_i(\sigma^*)\) and rewrite the above equation as
\[\sigma^*_{is} = \frac{e_{i1s} - e_{i1t}}{\beta_{i1s} e_{i1s}} k_i(\sigma^*) + \frac{\beta_{i1t} e_{i1t}}{\beta_{i1s} e_{i1s}} \sigma^*_{it},\quad 1\leq s,t\leq \rho.\]
Summing on \(s\) from \(1\) to \(\rho\) gives
\[1 = k_i(\sigma^*)\left(\sum_{p = 1}^\rho \frac{e_{i1p} - e_{i1t}}{\beta_{i1p} e_{i1p}}\right) + \sigma^*_{it}\left(\sum_{p = 1}^\rho \frac{\beta_{i1t} e_{i1t}}{\beta_{i1s} e_{i1s}}\right).\]
Setting 
\[E_{it} = \sum_{p = 1}^\rho \frac{e_{i1p} - e_{i1t}}{\beta_{i1p} e_{i1p}},\quad B_{it} = \sum_{p = 1}^\rho \frac{\beta_{i1t} e_{i1t}}{\beta_{i1s} e_{i1s}}\]
we may write
\[k_i(\sigma^*) = \frac{1 - \sigma_{it} B_{it}}{E_{it}}\]
and hence
\[\frac{1 - \sigma^*_{is} B_{is}}{E_{is}} = \frac{1 - \sigma^*_{it} B_{it}}{E_{it}},\quad 1\leq s,t\leq \rho.\]
It follows that
\[\sigma^*_{is} = \frac{E_{it} - E_{is}}{E_{it} B_{is}} + \frac{E_{is} B_{it}}{E_{it} B_{is}} \sigma^*_{it}.\]
So we may write each \(\sigma^*_{is}\) in terms of \(\sigma^*_{i1}\) and appropriate constants.

To write the relations in a more compact form, let
\[a_{is} = \frac{E_{i1} - E_{is}}{E_{i1} B_{is}}, \qquad b_{is} = \frac{E_{is} B_{i1}}{E_{i1} B_{is}},\]
hence we may write $\sigma^*_{is} = a_{is} + b_{is}\sigma^*_{i1}$ and
\begin{align*}
\sum_{r=1}^n \sum_{p=1}^\rho e_{r1p} \sigma^*_{rp} &
=\sum_{r=1}^n \sum_{p=1}^\rho \big(a_{rp} e_{r1p} + b_{rp} e_{r1p} \sigma^*_{r1}\big) \\
&
=\left(\sum_{r=1}^n \sum_{p=1}^\rho a_{rp} e_{r1p}\right) 
+ 
\sum_{r=1}^n \left(\sum_{p=1}^\rho b_{rp} e_{r1p}\right) \sigma^*_{r1} \\
& = a + \sum_{r=1}^n b_r\sigma^*_{r1},
\end{align*}
where 
\[a = \sum_{r=1}^n \sum_{p=1}^\rho a_{rp} e_{r1p},\qquad b_r = \sum_{p=1}^\rho b_{rp} e_{r1p}, \quad 1\leq r\leq n,\]
so 
\begin{align*}
u_i(\sigma^*_i, \sigma^*_{-i}) &= \frac{\sum_{p=1}^\rho \beta_{i1p} Q_1 e_{i1p} {\sigma^*_{ip}}^2}{a + \sum_{r=1}^n b_r\sigma^*_{r1}} \\
&= \frac{\sum_{p=1}^\rho \beta_{i1p} Q_1 e_{i1p} (a_{ip} + b_{ip}\sigma^*_{i1})^2}{a + \sum_{r=1}^n b_r\sigma^*_{r1}}.
\end{align*}
Now, for each $i\in I$, we define function $v_i$ from $[0, 1]^n$ to $\mathbb{R}$ by
\[
v_i(\tau_1,\ldots,\tau_n) = \frac{\sum_{p=1}^\rho \beta_{i1p} Q_1 e_{i1p} (a_{ip} + b_{ip}\tau_i)^2}{a + \sum_{r=1}^n b_r\tau_r}.
\]
Since $(\sigma^*_{11}, \ldots, \sigma^*_{n1})$ locally maximizes $u_i$, we may find it with usual techniques for maximizing $v_i$. As each $\sigma^*_{is}$ can be expressed in terms of $\sigma^*_{i1}$, this gives the whole point $\sigma^*$.  

%----------------------------------------------------------------------------

\section{Conclusion}\label{conclusion}
This research captures the competition among firms in a market where firms have to decide on which subset of products to produce  with differentiated
products. The object for the firms is to maximize their expected shared surplus. The competition among firms is modeled by a non-cooperative game, called PPM game, where  the utilities of agents is measured by their expected shared surplus.
The  Lagrange multipliers method is used  to compute an interior  Nash equilibrium (if there exist any) for a single market segmentation.

Future studies can focus on analyzing Nash equilibrium to predict strategic behavior of firms in more general markets with different markets.   The  solution concept of $\varepsilon$-Nash equilibrium provides the permission of the unilateral deviation of $\varepsilon$ value which may be useful for using  Lagrange multipliers method for finding a mixed Nash equilibrium of a PPM game in which supports of some mixed strategies in equilibrium are singletons.

% Acknowledgments here
\ACKNOWLEDGMENT{%
This research was partly established during the first author's sabbatical visit at Maastricht University. The authors wish to express their gratitude to Professor Dries Vermeulen, who without his support and guidance this study would not have been possible.
% Enter the text of acknowledgments here
}% Leave this (end of acknowledgment)

% Appendix here
% Options are (1) APPENDIX (with or without general title) or 
%             (2) APPENDICES (if it has more than one unrelated sections)
% Outcomment the appropriate case if necessary
%
% \begin{APPENDIX}{<Title of the Appendix>}
% \end{APPENDIX}
%
%   or 
%
% \begin{APPENDICES}
% \section{<Title of Section A>}
% \section{<Title of Section B>}
% etc
% \end{APPENDICES}

% References here (outcomment the appropriate case) 

% CASE 1: BiBTeX used to constantly update the references 
%   (while the paper is being written).
\bibliographystyle{informs2014} % outcomment this and next line in Case 1
\bibliography{ppm-references} % if more than one, comma separated

% CASE 2: BiBTeX used to generate mypaper.bbl (to be further fine tuned)
%\input{mypaper.bbl} % outcomment this line in Case 2

\end{document}